\documentclass[12pt]{article}
\pdfoutput=1

\usepackage{amsmath}
\usepackage{amssymb}
\usepackage{subfigure}
\usepackage{bm}   
\usepackage{verbatim} 
\usepackage[square, numbers]{natbib}

\usepackage{authblk}

\usepackage{latexsym,amsfonts,amsthm,amsbsy,multirow,textcomp,hyperref,wrapfig,datetime,cancel,subfigure,comment}

\usepackage{hyperref}
\hypersetup{colorlinks,
 linkcolor = blue,
 anchorcolor = red,
 citecolor = blue,
 filecolor = red,
 urlcolor = red,
 linktocpage}
 
\newcommand{\be}{\begin{equation}}
\newcommand{\ee}{\end{equation}}
\newcommand{\bea}{\begin{eqnarray}}
\newcommand{\eea}{\end{eqnarray}}

\DeclareSymbolFont{AMSb}{U}{msb}{m}{n}
\DeclareMathSymbol{\N}{\mathbin}{AMSb}{"4E}
\DeclareMathSymbol{\Z}{\mathbin}{AMSb}{"5A}
\DeclareMathSymbol{\R}{\mathbin}{AMSb}{"52}
\DeclareMathSymbol{\Q}{\mathbin}{AMSb}{"51}
\DeclareMathSymbol{\p}{\mathbin}{AMSb}{"50}
\DeclareMathSymbol{\I}{\mathbin}{AMSb}{"49}
\DeclareMathSymbol{\C}{\mathbin}{AMSb}{"43}
\DeclareMathSymbol{\D}{\mathord}{symbols}{"72}
\DeclareMathSymbol{\HH}{\mathbin}{AMSb}{"48}

\def\D{\nabla}

\def\({\begin{equation}}
\def\){\end{equation}}
\def\#{\begin{enumerate}}
\def\##{\end{enumerate}}

\newcommand{\ket}[1]{\left| #1\right>}
\newcommand{\bra}[1]{\left< #1\right|}

\def\phi{\varphi}

\begin{document}

\title{Remarks on Rindler Quantization}

\author{Ben Michel}\date{}
\affil{Department of Physics, University of California, Santa Barbara, CA 93106}

\maketitle
\begin{abstract}
\noindent
We review the quantization of scalar and gauge fields using Rindler coordinates with an emphasis on the physics of the Rindler horizon. In the thermal state at the Unruh temperature, correlators match their Minkowski vacuum values and the renormalized stress tensor vanishes, while at any other temperature the renormalized stress-energy diverges on the horizon. After giving a new derivation of some of these results using canonical quantization in the thermofield double state, we comment on the relevance of fluxes and boundary conditions at the horizon, which have arisen in calculations of entanglement entropy.

\end{abstract}

\section{Introduction}

One peculiar feature of relativistic quantum theories is the freedom to quantize in any time coordinate. This leads to interesting physics even in flat space, such as the Unruh effect \cite{Unruh:1976db}: accelerating observers quantize in Rindler time $\tau$ rather than Minkowski time $t$ and experience the Minkowski vacuum as a thermal state, a fact encoded in the Bogoliubov transformations between the accelerated and inertial mode functions.

Quantization in Rindler modes was first described by Fulling in his 1972 PhD. thesis \cite{Fulling:1972md} and is most often used as a tool to understand the thermal nature of the Minkowski vacuum \cite{Unruh:1976db,Israel:1976ur}. However it is also a perfectly valid way to describe operators in either Rindler wedge, and like any other quantization can be used to compute correlation functions wherever the coordinates are well-defined. At the Unruh temperature these computations ought to reproduce Minkowski expectation values, while at the physics at other temperatures can be quite different.

Correlation functions in Rindler quantization were the subject of intensive study in the 1970s and 1980s. However, the connection to Minkowski quantization is somewhat hidden in the literature, and has not to our knowledge been discussed in the context of the emerging links between entanglement and spacetime. It is the purpose of this note to highlight these old results, offer a new and more explicit derivation using the thermofield double state and explore the connection to issues relevant to entanglement entropy and the black hole information paradox.

Boulware was the first to show equivalence between the quantization of a free scalar in the Minkowski vacuum and Rindler quantization in the thermal state at the Unruh temperature, which is often called the Hartle-Hawking state \cite{Hartle:1983ai}. The argument is in appendix B of \cite{Boulware:1974dm}: an integral representation of the Hartle-Hawking two-point function can be manipulated into an integral representation of the Minkowski vacuum two-point function, using integral representations for the concomitant modified Bessel functions and an analytic continuation. This is enough to conclude equivalence.

More general thermal states are also interesting, and away from the Unruh temperature they are singular on the horizon. The thermal two-point function was computed by Dowker in $d=4$ \cite{Dowker:1977zj,Dowker:1978aza}. His main tool, Wick rotation, has played a prominent role in studies of Rindler space: at inverse temperature $\beta$ the Wick rotation of Rindler space is a Euclidean cone of opening angle $\beta$, which is amenable to classical techniques. In \cite{Dowker:1977zj} Dowker used Carslaw's result from 1919 \cite{carslaw_1919} for the propagator on the infinite cover of the plane before imposing periodicity with the method of images to obtain a contour integral representation of the Euclidean two-point function, which can be evaluated \cite{Dowker:1978aza} and continued \cite{Moretti:1995fa} to obtain Lorentzian correlators.

The renormalized thermal stress-energy tensor can be computed either by conformal methods or by direct differentiation of Dowker's correlator; using the latter method, Brown and Ottewill \cite{Brown:1985ri} found that the boost energy density diverges at the horizon except when $\beta$ is the inverse of the Unruh temperature. Candelas and Deutsch \cite{Candelas:1977zza} described a similar effect when a brick-wall boundary condition is imposed just outside the horizon, even at the Unruh temperature. Candelas \cite{Candelas1980} also studied some of these questions in the Schwarzschild geometry.

While the results of this note largely recapitulate the literature, our methods are different and it is instructive in any case to review these calculations since they bear on questions that had not arisen when they were first performed. In \S~\ref{minkrind} we give a new proof of the equivalence between the Minkowski vacuum and the Hartle-Hawking state using canonical quantization in the thermofield double state via explicit Lorentzian computation of the Wightman function.  This evaluation makes manifest convergence issues which demand an $i\epsilon$ prescription in agreement with the corresponding Minkowski correlator. Our computation does not simply extend to arbitrary $\beta$ but we study $\beta\rightarrow\infty$, the Boulware vacuum, and reproduce the results of Dowker in a tractable coincidence limit. We find a straightforward way to extend our discussion to U(1) gauge theory, which simplifies previous work \cite{Higuchi1992,Moretti:1996zt,Moretti:1996ws}. In \S~\ref{nonhh} we discuss non-Hartle-Hawking states. After reviewing the behavior of the stress tensor at the horizon in \S~\ref{stresstensor}, following \cite{Brown:1985ri} and \cite{Candelas:1977zza}, we close in \S~\ref{discussion} by addressing questions that arise from calculations of entanglement entropy: the necessity of flux and boundary condition sums at the horizon.

\section{Minkowski correlations from Rindler quantization}
\label{minkrind}

The Hartle-Hawking state is the Minkowski vacuum \cite{Unruh:1976db,Israel:1976ur} and so Rindler correlation functions in the Hartle-Hawkingstate should reproduce their Minkowski vacuum expectation values. Our point of comparison will be the two-point (Wightman) function of a free massless scalar in the Minkowski vacuum of flat $d$-dimensional spacetime, which is \cite{Birrell:1982ix}
\begin{equation}
\label{minkwightman}
\bra{0} \phi(x^M,t) \phi({x'}^M,t')\ket{0} = \frac{\Gamma\left(\frac{d-2}{2}\right)}{4\pi^{d/2}} \frac{1}{\left[(x^M-{x'}^M)^2-(t-t'-i\epsilon)^2\right]^{(d-2)/2}}.
\end{equation}
The index $M = 1\dots d-1$ and $\epsilon > 0$. Rindler quantization in the Hartle-Hawking state must reproduce this result.

Rindler coordinates in the right (left) wedge $(\tau,z, \vec{x})$ are related to the Minkowski coordinates $(t,x^M)$ by
\be
\label{coordxform}
t=(-)z\sinh\alpha\tau,\quad x^1=(-)z\cosh\alpha\tau,\quad \vec{x} = (x^2,\dots,x^{d-1}).
\ee
$z$ ranges from 0 to $\infty$ while the other coordinates are unbounded; with this choice $z$ is positive in each wedge. When written unadorned, $x$ refers to $\vec{x}$. We will set $\alpha\rightarrow 1$ for the remainder, but dependence on the acceleration can be restored by dimensional analysis.

The solutions to the Rindler equation of motion
\be
\label{rindeom}
(\partial_z^2 + z^{-1} \partial_z + \vec{\partial}^2 - z^{-2} \partial_{\tau}^2)\phi=0
\ee
involve the modified Bessel functions $I(z)$ and $K(z)$, the latter of which is finite as $z\rightarrow\infty$.\footnote{In the past and future wedges the argument becomes complex, and the relevant solutions involve the second Hankel function $H^{(2)}$ in place of $K$.} In the right and left wedges respectively, the mode functions 
\begin{equation}
f_{\omega k}(z,\tau,x) =K_{i\omega}(|k|z) e^{i(kx-\omega \tau)},\quad \quad\quad  \tilde{f}_{\omega k}(z,\tau,x) = K_{i\omega}(|k|z) e^{i(kx+\omega \tau)}
\end{equation}
parametrize the solutions to \eqref{rindeom} with the appropriate asymptotics. Solutions with $\omega>0$ are positive-frequency modes with respect to the generator of Rindler time translations in each wedge and accompany the creation/annihilation operators of a field expanded in Rindler coordinates just as planar mode functions accompany the Cartesian field expansion in flat space:
\begin{equation}
\label{fieldexpansion}
\phi^R(z,\tau,x)=\int_0^{\infty} d\omega \int d^{d-2}k\ N_{\omega k}\left[ f_{\omega k} (z,\tau,x) a^R_{\omega k} + \mbox{h.c.}\right]
\end{equation}
in the right wedge, and
\begin{equation}
\phi^L(z,\tau,x)=\int_0^{\infty} d\omega \int d^{d-2}k\ N_{\omega k}\left[ \tilde{f}_{\omega k} (z,\tau,x) a^L_{\omega k} + \mbox{h.c.}\right]
\end{equation}
in the left. The range of the $\omega$ integral reflects the division into positive and negative frequency modes. With the normalization
\be
N_{\omega k}^2 = (2\pi)^{d-2}\frac{\sinh \pi\omega}{\pi^2}
\ee
the modefunctions are Klein-Gordon orthonormal (which can be shown using the orthogonality relation $\int_0^\infty \frac{dx}{x} K_{i\omega}(x) K_{i\omega'}(x) = \frac{\pi^2}{2\omega\sinh\pi\omega}\delta_{\omega\omega'}$ -- proved quite recently, see \cite{Szmytkowski2010}) and
\be
[a^{i}_{\omega k},{a^{j}}^{\dagger}_{\omega' k'}] = \delta^{ij} \delta_{\omega \omega'}\delta_{kk'}
\ee
follows from the canonical commutation relation for $\phi$. Here $i,j$ are either L or R.

We will compute correlation functions in the thermofield double state
\(
\label{tfd}
|\text{TFD},\beta\rangle = \bigotimes_{\omega, k} Z_{\omega, k} \sum_{n} e^{-\frac{\beta
 n\omega}{2}} |\Theta(n,\omega,k)\rangle_{L}\ |n,\omega,k\rangle_{R}.
\)
$\Theta$ is a CPT conjugation picked out by the path integral preparation, with its P operator acting only on the $(\tau, z)$ plane:
\be
|\Theta(n, \omega,k) \rangle= |n, \omega,-k\rangle,
\ee
and the $Z_{\omega,k}$ are chosen such that $\bra{\text{TFD},\beta}\text{TFD},\beta\rangle=1$. The Hartle-Hawking state $\ket{\Omega}$ is the thermofield state \eqref{tfd} at $\beta=2\pi$ \cite{Unruh:1976db,Israel:1976ur},
\be
\label{hh}
\ket{\Omega}=\ket{\text{TFD},2\pi}.
\ee
The details of the computation depend on whether or not the operators are on the same side of the horizon. First we study one-sided correlators, putting both operators in the right wedge. It will be useful to separate the product of fields into the commutator and the anticommutator:
\begin{equation}
\phi(X)\phi(X') = \frac{1}{2}[\phi(X),\phi(X')] + \frac{1}{2}\{\phi(X),\phi(X')\}.
\end{equation}
The commutator is independent of the state and so the interesting piece is the anticommutator, a hermitean operator and the subject of our first computation. We have denoted the coordinates collectively by $X$ and will also abbreviate $(\omega,\vec{k})\equiv S$, $\int_0^{\infty} d\omega \int d^{d-2}k \equiv \int dS$. Taking the expectation value in the Hartle-Hawking state \eqref{hh} and using the field expansion \eqref{fieldexpansion}, one finds
\bea
\bra{\Omega} \{\phi^R(X), \phi^R(X') \}\ket{\Omega} = &\bra{\Omega} \left[\int dS dS' N_S N_S'\left(a_S f_S(X) + a^\dagger_S f^*_S(X)\right) \left(a_{S'} f_{S'}(X') + a^\dagger_{S'} f^*_{S'} (X')\right)\right]\ket{\Omega}\nonumber\\
&+ (X\leftrightarrow X').
\eea
The $aa$ and $a^{\dagger}a^{\dagger}$ terms do not contribute because the left and right occupation numbers must match, while the $a^\dagger a$ and $aa^\dagger$ terms can be computed using the properties of the Bose-Einstein distribution, since $\ket{\Omega}$ is a thermal state in $\omega$ with $\beta=2\pi$:
\(
\bra{\Omega} a^\dagger_{k\omega} a_{k'\omega'}\ket{\Omega}  = \frac{1}{e^{2\pi\omega}-1} \delta_{SS'}, \quad \bra{\Omega} a_{k\omega} a^\dagger_{k'\omega'}\ket{\Omega}  = \frac{e^{2\pi\omega}}{e^{2\pi\omega}-1} \delta_{SS'}.
\)
Their sum is $\coth\pi\omega$. After a short computation one finds
\begin{align}
\label{scalarint}
\bra{\Omega} \{\phi^R(X),\phi^R(X') \}\ket{\Omega} &= \int dS \coth\pi\omega\ N_{\omega k}^2\left[f_S(X) f^*_S(X') + f^*_S(X) f_S(X') \right]\nonumber\\
&= \int_0^\infty d\omega \int \frac{d^{d-2}k}{(2\pi)^{d-2}} \frac{\cosh \pi\omega}{\pi^2}\left[ e^{i[k(x-x')-\omega(\tau-\tau')]} K_{i\omega}(|k|z) K^*_{i\omega}(|k|z') + (X\leftrightarrow X')\right].
\end{align}
The evaluation of this integral is left to the appendix. The basic idea is to first do the integral without the cosh, using integral representations for the Bessel functions and the fact that the integrand is even in $\omega$. The cosh can then be accomodated via analytic continuation, which is only well-defined if we implement the $i\epsilon$ prescription below. The result is
\begin{align}
\label{tfdanticomm}
\bra{\Omega} \{\phi^R(X), \phi^R(X') \}\ket{\Omega} = &\frac{\Gamma\left(\frac{d-2}{2}\right)}{4\pi^{d/2}}\left[ \frac{1}{\left[z^2+z'^2-2zz'\cosh(\tau-\tau'-i\epsilon)+(\vec{x}-\vec{x}')^2\right]^{(d-2)/2}}\right.\nonumber\\ &\left.+\frac{1}{\left[z^2+z'^2-2zz'\cosh(\tau-\tau'+i\epsilon)+(\vec{x}-\vec{x}')^2\right]^{(d-2)/2}} \right].
\end{align}
The computation of the commutator yields a similar expression:
\begin{align}
[\phi^R(X), \phi^R(X')] = &\frac{\Gamma\left(\frac{d-2}{2}\right)}{4\pi^{d/2}}\left[ \frac{1}{\left[z^2+z'^2-2zz'\cosh(\tau-\tau'-i\epsilon)+(\vec{x}-\vec{x}')^2\right]^{(d-2)/2}}\right.\nonumber\\ &\left.-\frac{1}{\left[z^2+z'^2-2zz'\cosh(\tau-\tau'+i\epsilon)+(\vec{x}-\vec{x}')^2\right]^{(d-2)/2}} \right].
\end{align}
This vanishes except on the light cone. Together they imply the one-sided Wightman function
\be
\label{tfdwightman}
\bra{\Omega} \phi^R(X)\phi^R(X') \ket{\Omega} = \frac{\Gamma\left(\frac{d-2}{2}\right)}{4\pi^{d/2}}\frac{1}{\left[z^2+z'^2-2zz'\cosh(\tau-\tau'-i\epsilon)+(\vec{x}-\vec{x}')^2\right]^{(d-2)/2}}
\ee 
which is just the Minkowski expression \eqref{minkwightman} after the change of coordinates \eqref{coordxform}. The $\epsilon$s in \eqref{tfdwightman} and \eqref{minkwightman} are related by a $z$-dependent redefinition, but since $z>0$ this does not affect the sign.

The computation of the two-sided correlator is similar but the matrix elements take more work, again left to the appendix. The correlator
\be
\bra{\Omega} \phi^R(X)\phi^L(X') \ket{\Omega} = \frac{\Gamma\left(\frac{d-2}{2}\right)}{4\pi^{d/2}}\frac{1}{\left[z^2+z'^2+2zz'\cosh(\tau-\tau')+(\vec{x}-\vec{x}')^2\right]^{(d-2)/2}}
\ee 
is once more the Minkowski result \eqref{minkwightman} with the change of coordinates \eqref{coordxform}, after accounting for the extra sign in the transformation to the left wedge (recall we defined $z$ to be positive in both wedges). As explained in the appendix there is no need for an $i\epsilon$ prescription, which is consistent because points in different wedges are spacelike separated. The two-sided commutator vanishes, as required by causality:
\be
[\phi^R(X), \phi^L(X')]= 0.
\ee

These results extend to gauge theory. As reviewed in the appendix, the only difference between the scalar and gauge calculations is the presence of a polarization sum. The thermal trace includes a sum over longitudinal modes, but the same modes appear in the Minkowski calculation and drop out of gauge-invariant correlators. In Feynman gauge the propagator is just the scalar propagator times the metric and so
\be
\label{gaugeresult}
\bra{\Omega} A_{\mu}(X)A_{\nu}(X') \ket{\Omega} = g_{\mu\nu}\bra{\Omega} \phi(X)\phi(X') \ket{\Omega}.
\ee 
It follows that correlation functions of gauge invariant (and in Feynman gauge, gauge variant) operators match their Minkowski vacuum expectation values.

This correlation structure implies the familiar fact that the Rindler horizon is invisible in the Hartle-Hawking state, since correlators match their Minkowski vacuum expectation values. All trans-horizon probes necessarily are well-behaved: for example, the two-point function changes smoothly as an operator is dragged across the horizon, and the expectation values of all operators everywhere assume their Minkowski vacuum values.

\section{Non-Hartle-Hawking states}
\label{nonhh}

In other thermal states $\ket{\text{TFD},\beta}$ the Rindler horizon is far from invisible.
The anticommutator in the thermofield state \eqref{tfd} is
\begin{align}
\label{generalbeta}
\bra{\text{TFD},\beta} \{\phi^R(X), \phi^R(X') \}\ket{\text{TFD},\beta} = \int& dS \coth\left(\beta\omega/2\right) \frac{\sinh\left(\pi \omega\right)}{ \pi^2} e^{i(k\Delta x-\omega\Delta \tau)} K_{i\omega}(|k|z) K^*_{i\omega}(|k|z')\nonumber\\
& + (X\leftrightarrow X').
\end{align}
At $\beta\neq 2\pi$ we will not obtain Minkowski vacuum expectation values. The simplest state with $\beta\neq 2\pi$ is the Boulware vacuum \cite{Boulware:1974dm}
\be
\label{boulwarevac}
\ket{B}\equiv \ket{\text{TFD},\infty}=\ket{0}_L\otimes\ket{0}_R
\ee
obtained by taking $\beta\rightarrow \infty$ in \eqref{tfd}. It is annihilated by the $a^i_{\omega k}$. As in any product state there are no trans-horizon correlations, and their absence necessarily implies a firewall.

The argument is straightforward. The renormalized stress tensor $\overline{T}_{\mu\nu}$ is defined by subtracting off the Minkowski vacuum expectation value $\bra{0}T_{\mu\nu}\ket{0}$, which is divergent everywhere due to the short-distance correlation structure. When all trans-horizon correlators are zero,
\be
\bra{B} \overline{T}_{\mu\nu}(z=0,\tau,x)\ket{B} = -\bra{0} T_{\mu\nu}(z=0,\tau,x)\ket{0},
\ee
since $\bra{B}T_{\mu\nu}(z=0,\tau,x)\ket{B}=0$ follows immediately from the lack of trans-horizon correlations and the point-splitting \cite{Birrell:1982ix} definition of $T_{\mu\nu}$. The renormalized stress tensor at the horizon is therefore proportionate to the singular unrenormalized Minkowski vacuum expectation value.

Actually, any non-Hartle-Hawking thermal state is singular on the horizon \cite{Brown:1985ri} despite the thermal entanglement (albeit at the wrong temperature). Dowker computed the two-point function at arbitrary inverse temperature in $d=4$ using Euclidean techniques \cite{Dowker:1978aza}:
\be
\label{eq:dowker}
\langle \phi(z,\tau,x)\phi(z',0,0)\rangle_\beta = \frac{i}{4\pi\beta z z' \sinh\gamma}\frac{\sinh\left(2\pi\gamma/\beta\right)}{\cosh\left(2\pi\gamma/\beta\right)-\cosh\left(2\pi\tau/\beta\right)}
\ee
where
\be
\cosh\gamma = \frac{z^2+z'^2+| x|^2}{2zz'}.
\ee
As we review in the next section, these states have divergent stress-energy as $z\rightarrow 0$ whenever $\beta\neq 2\pi$. On the other hand, taking $\beta=2\pi$ and doing the appropriate Lorentzian continuation, it is easy to see that \eqref{eq:dowker} agrees with the Hartle-Hawking result \eqref{tfdwightman}. This is our point of comparison for the general thermal calculation in $d=4$.

The $\beta\rightarrow \infty$ limit of \eqref{generalbeta} is harder to compute since the integrand is odd in $\omega$, so the tricks applied to the Hartle-Hawking state in the previous section will not work here.\footnote{The integral representation $K_{i\omega}(|k|z)K^*_{i\omega}(|k|z') = \frac{1}{2} \int_{-\infty}^{\infty} d\lambda e^{i\omega\lambda} K_0(|k|\upsilon)$, where $\upsilon = z^2+{z'}^2+2zz'\cosh\lambda$, might be useful; finite $\beta$ and general $d$ are likely tractable.} However it is relatively easy to evaluate the integrals in the limit of coincident $\vec{x}$ in $d=4$, where we obtain 
\be
\label{eq:boulware}
\bra{B} \phi^R(z,\tau,x)\phi^R(z',\tau',x) \ket{B} =\frac{1}{4\pi^2(z^2-{z'}^2)}\left[\frac{1}{\Delta\tau+\log\frac{z}{z'}-i\epsilon} - \frac{1}{\Delta\tau-\log\frac{z}{z'}-i\epsilon}\right].
\ee 
This agrees with Dowker's result \eqref{eq:dowker} when ${x} = 0$ (i.e. $\gamma = \log(z/z')$). As either of the operators is taken to the horizon ($z$ or $z'\rightarrow 0$), \eqref{eq:boulware} vanishes. By contrast the Wightman function in the Hartle-Hawking state \eqref{tfdwightman} approaches $\frac{1}{z^2}$ when $z'\rightarrow 0$ at coincident ${x}$, which is just a reflection of the fact that all points on a boost orbit have the same distance from the origin.

\section{Stress-energy at the horizon}
\label{stresstensor}

Since the Wightman function in a general state in the thermofield double differs from the Wightman function in the Minkowski vacuum, it corresponds to a non-Minkowski vacuum distribution of stress-energy. The stress tensor can be computed either directly from \eqref{eq:dowker} \cite{Dowker:1987mn}, or by computing the response of the effective action to a change in the metric. Brown and Ottewill \cite{Brown:1985ri} took the latter approach in $d=4$ and computed the stress tensor using the dimensionally regularized effective action; in this scheme one obtains a vanishing Minkowski vacuum stress tensor and so the method agrees with the canonical computation in the vacuum subtraction scheme. When the manifold is conformally related to one on which the trace of the stress tensor vanishes, such as Rindler to Minkowski, the stress tensor on the original manifold is determined in terms of the $(a,c)$ anomalies of the theory and geometric data of the conformal relation. They find for the thermal expectation value of the stress tensor in the Rindler theory at inverse temperature $\beta$
\be
\label{eq:tmnu}
\bra{\text{TFD},\beta}\overline{T}_{\mu\nu}(z,\tau,x)\ket{\text{TFD},\beta} = \frac{1}{1440\pi^2 z^4}\left[\left(\frac{2\pi}{\beta}\right)^4-1\right]\left(g_{\mu\nu} + 4v_\mu v_\nu\right)
\ee
where $v_\mu=(z,\vec{0})$ is a unit vector proportionate to the boost Killing vector $\partial_\tau$. This is a renormalized stress tensor (it vanishes in the Minkowski vacuum) and a zero-temperature term has been omitted \cite{Brown:1985ri}. From \eqref{eq:tmnu} it is clear that the stress-energy diverges at the Rindler horizon at any non-Unruh temperature: there is a firewall. The states with $\beta\neq 2\pi$ are still thermally entangled and have two-sided correlations at finite $\beta$ but correspond to Euclidean path integrals with a conical deficit inserted at the origin on the $t=0$ slice.

It is interesting to contrast this behavior with the stress tensor obtained from a brick-wall quantization of the scalar field, where a boundary condition such as
\be
\phi(z=z_0, \tau,\vec{x})=0
\ee
is imposed on the ``stretched horizon'' at $z=z_0$ and the field expansion \eqref{fieldexpansion} modified to satisfy the boundary conditions. This quantization was studied in detail (in $d=4$) by Candelas and Deutsch \cite{Candelas:1977zza}. Letting $G_0$ denote the Hartle-Hawking Wightman function \eqref{tfdwightman}, they found
\be
D(X,X')=G_0(X,X') - \frac{i}{\pi}\int_0^{\infty} \frac{d\omega}{2\pi} e^{-i\omega\Delta \tau} \int\frac{d^2k}{(2\pi)^2} e^{ik\Delta x} \frac{K_{i\omega}(e^{i\pi} |k|z_0)}{K_{i\omega}(|k|z_0)} K_{i\omega}(|k|z) K_{i\omega}(|k|z')
\ee 
for the Wightman function in the Hartle-Hawking state with Dirichlet boundary conditions, and
\be
N(X,X')=G_0(X,X') + \frac{i}{\pi}\int_0^{\infty} \frac{d\omega}{2\pi} e^{-i\omega\Delta \tau} \int\frac{d^2k}{(2\pi)^2} e^{ik\Delta x} \frac{K'_{i\omega}(e^{i\pi} |k|z_0)}{K'_{i\omega}(|k|z_0)} K_{i\omega}(|k|z) K_{i\omega}(|k|z')
\ee 
with Neumann. The expectation value of the renormalized stress tensor is
\be
\bra{\text{TFD (brick wall)},2\pi} \overline{T}_{\mu\nu}(z,\tau,\vec{x}) \ket{\text{TFD (brick wall)},2\pi}= \text{diag}(c_1,c_2,c_3,c_3)
\ee
where $c_{1,2,3}$ depend on the choice of boundary conditions.\footnote{This expectation value is computed in the analog of the Hartle-Hawking state in the brick-wall quantization. The boundary condition imposes a quantization condition on the frequencies, and so the tensor product in \eqref{tfd} runs over a discrete set that depends on the boundary condition.} Candelas and Deutsch give integral representations for the $c_i$ and evaluate the stress tensor near the brick wall at $z= z_0$:
\be
c_1\sim -\frac{z^2}{360\pi^2 z_0(z-z_0)^3},\quad\quad c_2\sim \frac{(z-z_0)c_1}{2z_0 z^2},\quad\quad c_3 \sim \frac{c_1}{2z^2}
\ee
for both Dirichlet and Neumann conditions. The independence of the near-horizon stress tensor on the choice of boundary conditions is unexpected. They find a similar result for gauge fields. 

The Hartle-Hawking state with a brick wall is qualitatively similar to a thermal state at $\beta\neq 2\pi$, as both have divergent stress-energy at the horizon. However, the degree of divergence is different. The divergence structure of the brick-wall stress-energy follows from dimensional analysis and the fact that the stress tensor must reproduce the result for an unaccelerated barrier as $z_0\rightarrow \infty$.

\section{Discussion: Correlations vs. entanglement}
\label{discussion}

It is useful to contrast these results with calculations of entanglement entropy, which can be quite subtle \cite{Kabat:1995eq,Ohmori:2014eia}, especially in gauge theory \cite{Kabat:1995eq,Casini:2013rba,Donnelly:2014gva}. Gauge invariance implies \cite{Donnelly:2014gva} that the state that leads to the correct calculation of the entropy in ordinary Maxwell theory is the extended thermofield state
\be
\label{gaugetfdblock}
\ket{\text{eTFD},\beta}=\bigoplus_{E_\perp} \sqrt{p\left(E_\perp\right)} \bigotimes_{\lambda,\omega, k}Z_{\omega k}\sum_n e^{-\frac{\beta n\omega}{2}} |n,\lambda,\omega,-k;E_\perp\rangle_{L}\  |n,\lambda,\omega,k;E_\perp\rangle_{R}.
\ee
Here $E_\perp$ is a configuration of the normal components of the electric field on the horizon, $p\left(E_\perp\right)$ are a set of probabilities that were computed for $d=4$ Maxwell theory in \cite{Donnelly:2014fua,Donnelly:2015hxa}, and the polarizations $\lambda$ and an omitted ghost dressing are discussed around eq. \eqref{gtfd} in the appendix. The general structure of the state \eqref{gaugetfdblock} was described by Donnelly \cite{Donnelly:2014gva} in $d=2$, where $E_\perp$ is the only quantum number. The $n^\text{th}$ term in the sum \eqref{gaugetfdblock} is a product of states with $n$ photons on top of the coherent state $|0;E_\perp\rangle$
in which the normal electric field at the horizon is $E_\perp$.

The block-diagonal structure of \eqref{gaugetfdblock} is required by Gauss's law, which equates gauge-invariant operators at the boundary of the subregion with a sum of operators localized outside the subregion. Consequently, by causality the boundary operators must commute with any operators localized to the subregion, and the algebra of the subregion decomposes into superselection sectors: in the Rindler wedge they are labeled by the normal electric field $E_\perp$ at the horizon. The block-diagional structure in \eqref{gaugetfdblock} makes an additional Shannon contribution to the vacuum entanglement entropy of the Rindler wedge \cite{Donnelly:2014gva} which must be included in order to obtain a result consistent with conformal symmetry in $d=4$ Maxwell theory \cite{Donnelly:2015hxa}. In conformal theories the individual blocks are BCFT states, and their contribution to the entropy is a weighted sum of Affleck-Ludwig entropies \cite{Affleck1991,Ohmori:2014eia}.

Computations of correlation functions, or the stress tensor -- actual observables, unlike vacuum entanglement -- could in principle share this structure: it might have been necessary to do a flux sum in order to obtain the correct value. However, we computed eq. \eqref{gaugeresult}, which demonstrated the equivalence between the Rindler and Minkowski quantizations of Maxwell theory, using the naive thermofield state described in the appendix
\be
\label{gaugetfdnoblock}
\ket{\text{TFD},\beta}=\bigotimes_{\lambda,\omega, k}Z_{\omega k}\sum_n e^{-\frac{\beta n\omega}{2}} |n,\lambda,\omega,-k\rangle_{L}\  |n,\lambda,\omega,k\rangle_{R}
\ee
which has no flux sum. It is possible that the right computation in the block-diagonal state \eqref{gaugetfdblock} leads to \eqref{gaugeresult} as well, but the correct correlator can be obtained more simply using \eqref{gaugetfdnoblock} sans flux sum. In the scalar theory there is no gauge invariance to complicate the factorization of the Hilbert space, but the issue of boundary conditions at the horizon remains \cite{Ohmori:2014eia}. One might have considered a sum over horizon field values motivated by the vacuum path integral, but \eqref{tfdwightman} shows that correlators can be correctly computed without such a horizon sum.

\setcounter{secnumdepth}{0}
\section{Acknowledgements}
I am grateful to Will Donnelly, Don Marolf, Eric Mintun, Joe Polchinski and Mark Srednicki for extensive discussions. I especially thank Joe for suggesting the project, and Will and Mark for comments on the draft. I was supported by the NSF Graduate Research Fellowship Grant
DGE-1144085 and NSF Grant PHY13-16748 over the course of this work.

\appendix
\setcounter{secnumdepth}{0}
\section{Appendix}

This appendix describes the canonical quantization computation of the massless scalar two-point function in the Hartle-Hawking state. We also discuss the Boulware vacuum and the quantization of Maxwell theory.

Our ultimate goal is to evaluate the Wightman function but we will have to begin with the anticommutator:
\be
\label{appdxanticomm}
\bra{\Omega}\{\phi^R(X),\phi^R(X')\}\ket{\Omega} = \int_0^{\infty} \frac{d^{d-2}k}{(2\pi)^{d-2}} \frac{\cosh\pi\omega}{\pi^2}\left[ e^{i(k\Delta x-\omega \Delta\tau)} K_{i\omega}(|k|z)K_{i\omega}^*(|k|z') + (X\leftrightarrow X')\right].           
\ee
The first step is to evaluate the integral without the $\cosh\pi\omega$:
\begin{align}
\label{wocosh}
\frac{1}{\pi^2}\int_0^{\infty}& \int \frac{d^{d-2}k}{(2\pi)^{d-2}} \left[ e^{i(k\Delta x-\omega \Delta\tau)} K_{i\omega}(|k|z)K_{i\omega}^*(|k|z') + (X\leftrightarrow X')\right]\nonumber\\
&=\frac{1}{2\pi^2} \int_{-\infty}^{\infty} d\omega \int \frac{d^{d-2}k}{(2\pi)^{d-2}} \left[ e^{i(k\Delta x-\omega \Delta\tau)} K_{i\omega}(|k|z)K_{i\omega}^*(|k|z') + (X\leftrightarrow X')\right]\nonumber\\
&\equiv C(X,X') + (X\leftrightarrow X').
\end{align}
We proceed by using an integral representation of the modified Bessel function
\be
K_\alpha(t) = \frac{1}{2}(t/2)^{\alpha} \int_0^{\infty} \frac{du}{u^{\alpha+1}} e^{-u -\frac{t^2}{4u}}
\ee
and then doing the $\omega$ integral:
\begin{align}
\label{cxx}
C(X,X')&=\frac{1}{8\pi^2} \int_{-\infty}^{\infty} d\omega \int \frac{d^{d-2}k}{(2\pi)^{d-2}} e^{i(k\Delta x-\omega \Delta\tau)} \int_0^{\infty} \frac{du du'}{uu'}\left(\frac{zu'}{z'u}\right)^{i\omega} e^{-(u+u')} e^{-\frac{k^2}{4}\left(\frac{z^2}{u}+\frac{z'^2}{u'}\right)}\nonumber\\
&= \frac{1}{8\pi^2}\int \frac{d^{d-2}k}{(2\pi)^{d-3}}  e^{ik\Delta x} \int_0^{\infty} \frac{du du'}{uu'} e^{-(u+u')} e^{-\frac{k^2}{4}\left(\frac{z^2}{u}+\frac{z'^2}{u'}\right)} \delta\left(\log\frac{zu'}{z'u}-\Delta\tau\right)\nonumber\\
&= \frac{1}{8\pi^2} \int_0^{\infty} \frac{du}{u} e^{-u\left(1+\frac{z'}{z}e^{\Delta\tau}\right)}\int
 \frac{d^{d-2}k}{(2\pi)^{d-3}}\ e^{ik\Delta x} e^{-\frac{k^2
     z^2}{4u}\left(1+\frac{z'}{z} e^{-\Delta\tau}\right)}\nonumber \\ 
&= \frac{(2\pi)^{2-\frac{d}{2}}}{8\pi^2} \int_0^{\infty}
\frac{du}{u} e^{-u\left(1+\frac{z'}{z}e^{\Delta \tau}\right)} \left(\frac{z^2+zz'e^{-\Delta \tau}}{2u}\right)^{-\frac{(d-2)}{2}} e^{-u\left(\frac{x^2}{z^2+zz'e^{-\Delta \tau}}\right)}\nonumber\\
&= \frac{\Gamma\left(\frac{d-2}{2}\right)}{4\pi^{d/2}} \left(z^2+z'^2+2zz'\cosh\Delta\tau + \Delta x^2\right)^{-\frac{d-2}{2}}.
\end{align}
Note that the original integrand had to be even in $\omega$ in order for this method to work.

Now we return to the anticommutator, which differs from $\eqref{cxx}$ by a $\cosh\pi\omega$ factor. This factor can be almost be accomodated by analytically continuing the integral to $\Delta \tau\rightarrow \Delta\tau\pm i\pi$, but the integral diverges outside a strip in the complex $\Delta \tau$ plane and will require regulation. The problematic regions of the integral are at $\omega\rightarrow \pm \infty$. At large imaginary order, the modified Bessel functions have the expansion \cite{Dunster1990}
\be
K_{i\omega}(x) = -\left(\frac{\pi}{\omega\sinh\pi\omega}\right)^{1/2} \sin\left(\omega\log(x/2)-\gamma_\omega\right)+O(x^2),
\ee
where $\gamma_\omega$ is the phase of $\Gamma(1+i\omega)$. Therefore the integral \eqref{cxx} is only finite at $\omega\rightarrow\pm\infty$ if 
\be
\label{convergencebds}
-\pi < \mbox{Im }\Delta \tau < \pi.
\ee
Using symmetry in $\omega$, the anticommutator \eqref{appdxanticomm} is
\be
\bra{\Omega}\{\phi^R(X),\phi^R(X')\}\ket{\Omega} = \int_{-\infty}^{\infty} \frac{d^{d-2}k}{(2\pi)^{d-2}} \frac{e^{\pi\omega}+e^{-\pi\omega}}{4\pi^2}\left[ e^{i(k\Delta x-\omega \Delta\tau)} K_{i\omega}(|k|z)K_{i\omega}^*(|k|z') + (X\leftrightarrow X')\right].
\ee
The only effect of the cosh is to multiply the integrand in \eqref{cxx} by an $e^{\pm\pi\omega}$ factor but this renders the integral divergent. The integral can be regulated by inserting a factor $e^{\mp\epsilon \omega}$ into the integrand, the sign depending on whether the offending divergence is at $\pm \infty$; $\epsilon>0$ will be taken to zero at the end of the calculation. The integral with the $e^{\pm\pi\omega}$ factor and the necessary regulator $e^{\mp \epsilon \omega}$ can then be computed by continuing $\Delta \tau\rightarrow \Delta \tau \pm i\pi \mp i\epsilon$ in \eqref{cxx} ($\Delta \tau\rightarrow \Delta \tau \mp i\pi \pm i\epsilon$ in the $X\leftrightarrow X'$ term in \eqref{wocosh}). This procedure is well-defined since the continuation stays within the convergence bounds \eqref{convergencebds}. One evaluates 
\begin{align}
\int_{-\infty}^{\infty} \frac{d^{d-2}k}{(2\pi)^{d-2}} e^{\pi\omega}e^{i(k\Delta x-\omega \Delta\tau)} K_{i\omega}(|k|z)K_{i\omega}^*(|k|z')  &\rightarrow \int_{-\infty}^{\infty} \frac{d^{d-2}k}{(2\pi)^{d-2}} e^{\pi\omega} e^{-\omega\epsilon} e^{i(k\Delta x-\omega \Delta\tau)} K_{i\omega}(|k|z)K_{i\omega}^*(|k|z') \nonumber\\
&=\int_{-\infty}^{\infty} \frac{d^{d-2}k}{(2\pi)^{d-2}} e^{i(k\Delta x-\omega \Delta\tau)} K_{i\omega}(|k|z)K_{i\omega}^*(|k|z') \biggr\rvert_{\Delta\tau\rightarrow \Delta\tau+i\pi-i\epsilon}
\end{align}
and similarly for the other terms. Combined with the result \eqref{cxx} for the integral to be continued,
\begin{align}
\bra{\Omega} \{\phi^R(X), \phi^R(X') \}\ket{\Omega} = &\frac{\Gamma\left(\frac{d-2}{2}\right)}{4\pi^{d/2}}\left[ \frac{1}{\left[z^2+z'^2-2zz'\cosh(\tau-\tau'-i\epsilon)+(\vec{x}-\vec{x}')^2\right]^{(d-2)/2}}\right.\nonumber\\ &\left.+\frac{1}{\left[z^2+z'^2-2zz'\cosh(\tau-\tau'+i\epsilon)+(\vec{x}-\vec{x}')^2\right]^{(d-2)/2}} \right].
\end{align}
Now we compute the commutator. Using the mode expansion \eqref{fieldexpansion},
\be
\label{appdxcomm}
[\phi^R(X),\phi^R(X')] = \int_0^{\infty} \frac{d^{d-2}k}{(2\pi)^{d-2}} \frac{\sinh\pi\omega}{\pi^2}\left[ e^{i(k\Delta x-\omega \Delta\tau)} K_{i\omega}(|k|z)K_{i\omega}^*(|k|z') - (X\leftrightarrow X')\right].           .
\ee
This differs from \eqref{appdxanticomm} by the sinh instead of the cosh, and the relative sign on the $(X\leftrightarrow X')$ term. It follows that the integrand is still even in $\omega$ and so we can use the result \eqref{cxx}, obtaining
\begin{align}
[\phi^R(X), \phi^R(X')]= &\frac{\Gamma\left(\frac{d-2}{2}\right)}{4\pi^{d/2}}\left[ \frac{1}{\left[z^2+z'^2-2zz'\cosh(\tau-\tau'-i\epsilon)+(\vec{x}-\vec{x}')^2\right]^{(d-2)/2}}\right.\nonumber\\ &\left.-\frac{1}{\left[z^2+z'^2-2zz'\cosh(\tau-\tau'+i\epsilon)+(\vec{x}-\vec{x}')^2\right]^{(d-2)/2}} \right].
\end{align}
after doing the requisite regulations and analytic continuations as described above. From \eqref{appdxanticomm} and \eqref{appdxcomm} we obtain the Wightman function \eqref{tfdwightman} in the Hartle-Hawking state. We could not have computed the Wightman function directly using the methods of this appendix, since the integrand is not even in $\omega$.

Now we consider operators inserted on opposite sides of the Rindler horizon. The anticommutator in the Hartle-Hawking state is
\bea
\label{oppanticomm}
\bra{\Omega} \{\phi^R(X), \phi^L(X') \}\ket{\Omega} = &\bra{\Omega} \left[\int dS dS' N_S N_S'\left(a^R_S f_S(X) + {a^R}^\dagger_S f^*_S(X)\right) \left(a^L_{S'} \tilde{f}_{S'}(X') + {a^L}^\dagger_{S'} \tilde{f}^*_{S'}(X')\right)\right]\ket{\Omega}\nonumber\\
&+ (X\leftrightarrow X').
\eea
The $aa^\dagger$ and $a^{\dagger}a$ terms do not contribute because the left and right boson numbers must match, while the $a a$ and $a^\dagger a^\dagger$ terms can be computed from the thermofield double:
\begin{align}
\label{aatfd}
\bra{\Omega} a^R_{\omega k} a^L_{\omega' k'}\ket{\Omega}  &= \frac{1}{1-e^{-\beta\omega/2}}\sum_{n,n'} \bra{n',\omega,k}_R\bra{n',\omega,-k}_L e^{-\frac{\beta\omega}{2}(n+n')} a^R_{k\omega} a^L_{k'\omega'}\ket{n,\omega,-k}_L\ket{n,\omega,k}_R\nonumber\\
&= \frac{\delta_{\omega\omega'}\delta_{k(-k')}}{1-e^{-\beta\omega/2}}\sum_{n,n'} n\bra{n',\omega,k}_R\bra{n',\omega,-k}_L e^{-\frac{\beta\omega}{2}(n+n')} \ket{n-1,\omega,-k}_L\ket{n-1,\omega,k}_R\nonumber\\
&= \frac{1}{2}\ \text{csch} \left(\frac{\beta\omega}{2}\right)\delta_{\omega\omega'}\delta_{k(-k')}\nonumber\\
&=\left(\bra{\Omega} {a_{\omega k}^R}^\dagger {a_{\omega k}^L}^\dagger\ket{\Omega}\right)^\dagger.
\end{align}
The first equality is slightly nontrivial: inserting $a^R_{\omega k} a^L_{\omega' k'}$ into the thermofield state \eqref{tfd}, there are two tensor products over the momentum quantum numbers in the bra and the ket (call them $S_\text{bra}$ and $S_\text{ket}$) which have been omitted here. However, unless $\delta_{\omega\omega'}\delta_{k(-k')}=1$ there will be a mode number mismatch between the left and the right; the orthonormality of the $\ket{n,\omega,k}$ ensures that $S_\text{bra}=S_\text{ket}\equiv (\bar{\omega},\bar{k})$, and except for $\bar{\omega}=\omega$, the normalization factor $Z_{\bar{\omega}}= (1-e^{-\beta \bar{\omega}/2})^{-1}$ cancels against the thermal sum since there are no insertions with those quantum numbers. This implies the first equality of \eqref{aatfd}.

Plugging \eqref{aatfd} into \eqref{oppanticomm} one obtains for the anticommutator
\begin{align}
\bra{\Omega}\{\phi^R(X),\phi^L(X')\}\ket{\Omega} =\int_0^{\infty} \frac{d^{d-2}k}{(2\pi)^{d-2}} \frac{1}{\pi^2}\left[ e^{i(k\Delta x-\omega \Delta\tau)} K_{i\omega}(|k|z)K_{i\omega}^*(|k|z') + (X\leftrightarrow X')\right]
\end{align}
which is the integral without the cosh evaluated above. Thus there is no need for analytic continuation and the answer is just
\be
\bra{\Omega}\{\phi^R(X),\phi^L(X')\}\ket{\Omega}=\frac{\Gamma\left(\frac{d-2}{2}\right)}{4\pi^{d/2}} \left(z^2+z'^2+2zz'\cosh\Delta\tau + \Delta x^2\right)^{-\frac{d-2}{2}}.
\ee
Since we did not analytically continue there is no need to regulate the integrals with an $i\epsilon$ prescription, which is perfectly consistent with the Minkowski description since insertions in different wedges are necessarily at spacelike separation. This also implies that the commutator vanishes:
\be
[\phi^R(X),\phi^L(X')]=0
\ee
as required by causality.

Next we will attempt to compute the Wightman function in the Boulware vacuum $\ket{B}$. Using the expansion \eqref{fieldexpansion} and the fact that $a\ket{B}=0$, one finds for the anticommutator
\be
\bra{B}\{\phi^R(X),\phi^R(X')\}\ket{B} = \int_0^{\infty} \frac{d^{d-2}k}{(2\pi)^{d-2}} \frac{\sinh\pi\omega}{\pi^2}\left[ e^{i(k\Delta x-\omega \Delta\tau)} K_{i\omega}(|k|z)K_{i\omega}^*(|k|z') + (X\leftrightarrow X')\right].           
\ee
This integral cannot be evaluated using the methods of this appendix, since the integrand is odd in $\omega$. However, the computation simplifies at $x=x'$ in $d=4$:
\begin{align}
\bra{B}\phi^R(X)\phi^R(X')\ket{B} &= \int_0^{\infty} \frac{d^2k}{(2\pi)^2} \frac{\sinh\pi\omega}{\pi^2} e^{-i\omega \Delta\tau} K_{i\omega}(|k|z)K_{i\omega}^*(|k|z')\nonumber\\
&= \frac{1}{4\pi^3} \int_0^\infty d\omega \sinh\pi\omega\ e^{-i\omega\Delta\tau} \left(\frac{z}{z'}\right)^{i\omega} \int_0^{\infty} du\ du' ds \left(\frac{u'}{u}\right)^{i\omega} e^{-(u+u')}e^{-s(z^2u'+{z'}^2u)}\nonumber\\
&= \frac{1}{4\pi^2(z^2-{z'}^2)}\left[\frac{1}{\Delta\tau+\log\frac{z}{z'}-i\epsilon} - \frac{1}{\Delta\tau-\log\frac{z}{z'}-i\epsilon}\right] 
\end{align}
after some manipulation. Again the $i\epsilon$ prescription is obtained by demanding convergence. This agrees with Dowker's result \eqref{eq:dowker} when $\vec{x}=0$, and approaches $\left(z\Delta\tau\right)^{-2}$ as $z\rightarrow z'$; in the same limit the thermofield correlator \eqref{tfdwightman} approaches $\left(z(1-\cosh\Delta\tau)\right)^{-2}$.

Last, we study the Rindler quantization of abelian Maxwell theory. Our goal will be to show that correlation functions of gauge-invariant operators in the Hartle-Hawking state, such as
\be
\bra{\Omega} F_{\mu\nu}(X) F_{\rho\sigma}(X')\ket{\Omega},
\ee
agree with their Minkowski expectation values. We will do this by showing that the Wightman function 
\be
\bra{\Omega} A_\mu(X) A_\nu(X') \ket{\Omega}
\ee
agrees with the Minkowski vacuum Wightman function
\be
\bra{0} A_\mu(X) A_\nu(X') \ket{0}
\ee
in Feynman gauge.

Ignoring the ghosts, the Lagrangian in Feynman gauge is
\be
\mathcal{L} = -\frac{1}{2} \left(\partial_\mu A_\nu\right)^2
\ee
and so the quantization of the gauge field is just that of a $d$ scalars.  In the right wedge the $d$ scalar fields can be expanded as
\(
A^R_\mu = \sum_{\lambda=0}^{d-1}
\int_0^{\infty} d\omega \int d^{d-2}k\ N_{\omega k}\left[ f_{\omega k} (z,\tau,x) a^R_{\lambda, \omega k}\epsilon_{\mu,\omega
    k}^{\lambda} + \mbox{h.c.}\right]
\)
where the $\epsilon_{\mu,\omega
    k}^{\lambda}$ are polarization vectors and $\lambda$ runs over $d$ polarizations, including two unphysical. Choosing $N_{\omega k}$ as for the scalar gives
\be
[a_{\lambda,\omega k}, a^\dagger_{\lambda',\omega'k'}]
= g_{\lambda \lambda'}\delta_{\omega\omega'} \delta_{kk'}
\ee
if we require the $\epsilon$ to obey
\be
\epsilon_{\mu,\omega
    k}^{\lambda} (\epsilon_{\mu,\omega
    k}^{\lambda'})^* = g^{\lambda \lambda'}
\ee
and
\be
\label{feynmansum}
\sum_{\lambda} \epsilon_{\mu,\omega
    k}^{\lambda} (\epsilon_{\nu,\omega
    k}^{\lambda})^* = g_{\mu\nu}.
\ee

We take the gauge theory analog of the thermofield state \eqref{tfd} to be
\(
\label{gtfd}
|\text{TFD}\rangle = \bigotimes_{\lambda,\omega, k} Z_{\omega k} \sum_{n} e^{-\frac{\beta\omega n}{2}} |n,\omega, -k,\lambda\rangle_{L}\  |n,\omega, k,\lambda\rangle_{R}.
\)
Note the absence of a flux sum. The state \eqref{gtfd} lives in a Hilbert space that is not obviously physical, as it includes states of negative norm. However, such states are BRST-exact and thus equivalent to zero in the cohomology of physical states; they do not contribute to gauge-invariant correlation functions. In order to describe an actual state in the gauge-fixed Hilbert space \eqref{gtfd} must technically be dressed with a tensor factor describing the ghost fields, but in the abelian theory the ghosts decouple and their tensor factor does not even affect the correlation functions of gauge-variant operators.

As usual our object of interest will be the Wightman function in the state \eqref{gtfd} at $\beta=2\pi$. After a short computation entirely analogous to the scalar case, one finds the anticommutator
\begin{align}
\bra{\Omega} \{A^R_\mu(X),A^R_\nu(X') \}\ket{\Omega}= \int_0^\infty &d\omega \int \frac{d^{d-2}k}{(2\pi)^{d-2}} \frac{\cosh \pi\omega}{\pi^2}\left[ e^{i(k\Delta x-\omega\Delta\tau)} K_{i\omega}(|k|z) K^*_{i\omega}(|k|z')\sum_{\lambda} \epsilon^{\lambda}_{\mu,\omega k}(\epsilon^{\lambda}_{\nu,\omega k})^*  \right.\nonumber\\
&+ (X\leftrightarrow X')\Biggr].
\end{align}
This differs from the scalar anticommutator \eqref{scalarint} only by the presence of the polarization sum. The Feynman gauge expression \eqref{feynmansum} for the polarization sum leads immediately to the result \eqref{gaugeresult} and the conclusion that gauge-invariant correlation functions reproduce their Minkowski expectation values. The argument for trans-horizon correlators proceeds identically.

\bibliographystyle{toine}
\bibliography{references.bib}

\end{document}